\documentclass[aps,preprint,amssymb,12pt,floatfix]{revtex4}
\usepackage[pdftex]{graphicx}
\usepackage{lscape,graphicx}
\usepackage{rotating}
\usepackage{epstopdf}
\usepackage{color}
\usepackage{amsmath,amsthm}
\usepackage{wrapfig}

\begin{document}
\title{RNA under Tension: Folding Landscapes, Kinetic Partitioning Mechanism, and Molecular Tensegrity}
\author{Jong-Chin Lin$^1$, Changbong Hyeon$^2$, and D. Thirumalai$^1$}
\affiliation{$^1$Biophysics Program, Institute for Physical Science and Technology, University of Maryland, College Park, MD 20742, USA}
\affiliation{$^2$School of Computational Sciences, Korea Institute for Advanced Study, Seoul 130-722, Republic of Korea}
\date{\today}
\begin{abstract}
Non-coding RNA sequences play a great role in controlling a number of cellular functions, thus raising the need to understand their complex conformational dynamics in quantitative detail.  In this perspective, we first show that single molecule pulling experiments when combined with  with theory and simulations can be used to quantitatively explore the folding landscape of nucleic acid hairpins, and riboswitches with tertiary interactions. Applications to riboswitches, which are non-coding RNA elements that control gene expression by undergoing dynamical conformational changes in response to binding of metabolites, lead to an organization principle that assembly of RNA is determined by the stability of isolated helices.  We also point out the limitations of single molecule pulling experiments, with molecular extension as the only accessible parameter, in extracting key parameters of the folding landscapes of RNA molecules. 
\end{abstract}
\maketitle

Ever since the pioneering discovery that RNA molecules can act as enzymes an increasing repertoire of functions have been associated with these bewilderingly complex biological molecules \cite{DoudnaNature02}. Even the ribosome that helps in  protein synthesis can be legitimately considered to be a ribozyme (RNA enzyme) because the peptidyl transfer center, the site at which the peptide bond is formed, is devoid of proteins \cite{NissenScience00}. More recently, it has been discovered  that RNA interference, involving small interference RNA and micro RNA, plays a vital role in post transcriptional gene regulation \cite{Matzke05NatReviews}.  
The finding that  riboswitches \cite{Winkler05ARM,Montange08ARB}, a class of RNA molecules, can by themselves control gene expression, without the participation of proteins, further underscores the important role RNA plays in key cellular functions.  In a majority of these examples folding and conformational changes associated with RNA molecules are at the center stage.  

RNA folding is considerably more complicated than the better studied protein folding for a number of reasons \cite{Thirum05Biochem}. First, the building blocks of RNA (A, U, G, and C) are chemically  similar except for the size and shapes of the bases.  As a consequence, RNA molecules are closer to homopolymers than proteins. Second, only about 50\% of nucleotides in RNA form Watson-Crick base pairs (A-U and G-C) whereas the remaining nucleotides are in bulges, loops, and other architectures \cite{Dima05JMB}. Finally, the highly charged nature of the phosphate groups, which makes RNA a polyelectrolyte, implies that folding to a compact structure cannot take place without the presence of counterions \cite{Heilman-Miller01JMB}.  Some of these aspects are reflected in the stability gap (the difference in the free energies of the native  and low-lying structures) for RNA being  not as large as it is in proteins \cite{Thirum05Biochem,Guo92JCP}. Thus,  during the folding process RNA can readily adopt  alternate folds, which although makes possible the extraordinary range of functions that are associated with RNA, also makes the study of their folding complicated.  

The complexity of RNA folding is succinctly summarized using the Kinetic Partitioning Mechanism (KPM) according to which a pool of unfolded molecules partitions into two distinct populations \cite{Guo95BP} that reach the Native Basin of Attraction (NBA) by vastly different time scales under folding conditions.  A fraction, $\Phi$ of unfolded molecules, folds rapidly to the NBA while the remaining fraction ($1 - \Phi$) is kinetically trapped in multiple competing basins of attraction (CBAs) (Fig. 1a). Transitions from the CBAs to the NBA, which occur by partial or global unfolding of the conformations in the CBAs~\cite{Pan97JMB} could take minutes or longer as shown experimentally for the well studied {\it Tetrahymena} ribozyme with $\Phi \approx 0.1$ \cite{Pan97JMB,Zhuang00Science}.  More recently, temperature jump experiments of  RNA pseudoknots \cite{narayanan2011JACS}  show that the KPM quantitatively describes their folding. 

Although global folding of RNA molecules is well understood the details of the underlying folding landscape are only starting to emerge thanks in part to advances in single molecule pulling Laser Optical Tweezers (LOT) experiments \cite{Woodside08COCB}. 
In LOT experiments the ends of  RNA of interest are tethered to handles (DNA or hybrid DNA/RNA), which are themselves attached to spherical beads that are localized  by laser traps (Fig.~1b). 
Mechanical force can be transmitted to RNA by moving one of the beads at a constant loading rate, $r_f = k_{trap}v$, ($k_{trap}$ is the force constant associated with the harmonic trap potential and $v$ is the pulling velocity). In this mode, the experiments yield force as a function of  extension, $R$, of the RNA (Fig.~1c) referred to as force-extension curves (FECs)~\cite{Bustamante01Sci,Bustamante03Science}.  
The FECs could be used to infer the order in which the structural elements are ruptured as the force is increased~\cite{Bustamante03Science}. 
Alternatively, constant $f$, applied to the ends of RNA by a suitable feedback techniques, could be used to generate mechanical folding trajectories expressed, which yield, $R_{sys}(t)$ (Fig.~1b) projected along the force axis as a function of time, $t$ (Fig.~1d provides a sample trajectory for a DNA hairpin \cite{Block06Science}). 
In analyzing the $R_{sys}(t)$ data to obtain equilibrium free energy profile $F(R)$ as a function of $R$ (Fig.~1b for the difference between $R$ and $R_{sys}$) a few assumptions are made: (i) Transverse fluctuations at the applied forces are negligible. (ii) Dynamics of $R_{sys}(t)$ mirrors $R(t)$. (iii) The system of interest ergodically samples the conformational space on the observation time scale so that the extracted $F(R)$ from $R(t)$ by appropriate deconvolution methods are the equilibrium profiles. Despite the restriction that LOT experiments only provide  one dimensional $F(R)$ they have given insights into folding of RNA at the single molecule level. In this perspective we show that simulations based on coarse-grained models \cite{Hyeon11NatComm,Sim12COSB} of RNA hairpin and riboswitches, experimental data, and theoretical considerations can be combined to extract some key aspects of the folding landscapes and dynamics.   Our purpose here is not to merely compare theoretical predictions and experiments but rather show both the successes and ultimately the challenges that need to be overcome in order to realize the potential of the ideas, methods, and concepts sketched here. 
\\

{\bf Complexity of RNA hairpin formation.}
RNA hairpins are  the simplest but the most ubiquitous motifs that form the building blocks of higher order structures.   
A substantial energetic contribution to the folded states of RNA comes from base pairings, the majority of which participate in forming stem-loop structures, namely hairpins.  For over four decades considerable effort has been made to study the thermodynamics and kinetics of the folding of simple RNA hairpins by using ensemble measurements, initiated by temperature ($T$) jump \cite{CrothersBIOCHEM72,porschke1973BP,Porschke74BPC,CrothersJMB74,Bloomfield00Book}. These experiments showed that generically RNA hairpins fold in an approximate two-state manner with a time constant of about $10 \mu s$.  Recent high resolution ensemble experiments \cite{Ma06JACS,Ma07PNAS} and single-molecule pulling experiments \cite{Bustamante01Sci,Woodside06PNAS} have shown that hairpin formation could be more complex than previously thought.   
Single molecule force measurements of RNA hairpin dynamics using LOT, pioneered by Liphardt \emph{et al.} \cite{Bustamante01Sci},  showed that a single RNA hairpin undergoes reversible folding and unfolding transitions in a narrow range of $f$ values  ($13-15$ pN).  The trajectory measuring the time-dependent changes in the extension, $R(t)$, jumps predominantly between two values one corresponding to the folded state and the other to the unfolded state. To a first approximation, such folding trajectories could be analyzed using an apparent two-state model, just as in the classic $T$-jump experiments. 

A long single time trace of RNA hairpin generated in LOT experiments is sufficient to extract both the thermodynamic and kinetic features of hairpin dynamics as long as it is assured that the RNA molecule ergodically samples its conformational space \cite{Bustamante01Sci,Hyeon08PNAS}. 
By using the average dwell times of RNA in the folded (small values of $R(t)$) and unfolded states (large values of $R(t)$) as a function of $f$ the equilibrium constant between the  two states, $K_{eq}(f)=\tau_{U}(f)/\tau_{F}(f)$, has been directly measured.   Extrapolation of  the measurements to $f = 0$  showed that the extracted stability values of P5ab and P5abc, the core secondary elements of the P4-P6 domain of the \emph{Tetrahymena} ribozyme, are consistent with those from bulk measurement ($K_{eq}(f\rightarrow 0)\approx K_{eq}^o$) \cite{Bustamante01Sci}, thus reinforcing the two state model for folding thermodynamics of generic RNA hairpins. 
A more detailed view of RNA hairpin thermodynamics was provided using realistic simulations \cite{Hyeon05PNAS},  which obtained the equilibrium phase diagram of  a 22-nt RNA hairpin as a function of $f$ and $T$ (Fig.2a). 
The phase diagram also revealed that there are predominantly only two states although this study provided hints of fine structure in the phase diagram.  The two states are separated by a first order phase coexistence line expressed in terms of a set of critical points ($T_m$, $f_m$) \cite{Hyeon05PNAS,Hyeon08JACS}. Thus,  from a thermodynamic perspective it appears that the phase diagram of a RNA hairpin in the $T$ and  $f$ variables could be approximated as a two-state system.
\\

{\bf Where folding starts matters.} The two-state model for RNA  hairpins hides the complex dynamics of the  RNA hairpin, which have been recently revealed using realistic simulations by varying $f$ and $T$ \cite{Hyeon08JACS}, and by advances in experimental methods \cite{LaPorta11BJ}. 
 In addition, recent $T$-jump  kinetic experiments  showed hairpins form in multiple steps \cite{Ma06JACS,Ma07PNAS}, which challenges the conventional notion that small nucleic acid hairpins could be modeled using only two states.  The structural origin of the complexity is due to the link between the establishment of local base-pairs  and the global hairpin formation, which were clearly demonstrated using coarse-grained simulations \cite{Hyeon05PNAS,Hyeon06Structure} and detailed all atom models \cite{Garcia08JACS,Bowman08JACS}. Hyeon and Thirumalai showed that the folding landscape  hairpins formation requires at least two reaction coordinates \cite{Hyeon06BJ,Hyeon08JACS}. 
Besides  $R$,  a collective variable describing the average deviation of loop dihedral angles from the native value is needed. Two-dimensional free energy surface at $f=f_m\approx 0$ and $T=T_m$, calculated using these variables 
shows that  the folding landscape of even a simple RNA hairpin is rugged, (especially at high $T_m$ and low $f_m$) explaining the observed complex kinetics \cite{Ma06JACS,Ma07PNAS}.  

The complexity of the kinetics of RNA hairpin formation is evident when folding is initiated from different parts of the landscape, which can be achieved by preparing the initial conformations by $T$-jump or by using high stretching forces. Remarkably, the refolding pathways of hairpin formation from a fully stretched initial state upon $f$-quench are distinct from the folding pathways observed in $T$-quench refolding (Fig.2b). 
The initial conformations of RNA hairpin under high tension are fully stretched and are structurally homogeneous. The various conformations largely differ in the internal degrees of freedom while the overall end-to-end distance is large, resulting in substantial deviations of the conformation of the tetra loop from the native structure. 
Thus, the first step in the hairpin formation from the initially stretched conformations is the tetra-loop formation, corresponding to the slow nucleation ($I_{SL}^f$ state in Fig.~2b) stage. The high entropic cost to establish the correct loop dihedral angles  makes the loop formation dynamics  unusually slow. 
 Subsequent to the nucleation step the zipping of remaining base pairs leads to rapid hairpin formation.  Thus, hairpin forms by the classic mechanism (establishment of base pair contact near the loop followed by a zipping process) when folding is initiated by f-quench (Fig~2b). 
 
In contrast, upon $T$-quench, refolding commences from a broad thermal ensemble of unfolded conformations (Fig.~2b). As a result, nucleation can originate from regions other than near the tetra-loop ($I_{SL}^f$, $I_{SL}^T$, and $I_{LL}$ states in Fig.~2b). Consequently, the pathway diversity is greater when hairpin formation is initiated by $T$-quench rather than f-quench.  The differences in the folding mechanism between these two methods to trigger hairpin formation are entirely due to the variations in the initial conformations. Exploring the details of the heterogeneous kinetics requires multiple probes that control the conformations of the ensemble of unfolded states.   
 Our studies also showed  that the complexity of energy landscape observed in ribozyme experiments is already reflected in the formation of simple RNA hairpins.  
\\
 
{\bf Folding landscapes from pulling experiments.}
Besides yielding stability and hopping rates between various states, single-molecule pulling experiments have been used to obtain one-dimensional free energy profiles as a function of $R$. It should be noted that the only directly measurable quantity  in LOT experiments are the time dependent changes in the distance between the beads, $R_{sys}(t)$ (Fig.~1b), at a fixed $f$. 
What is of interest, however, is the free energy profile, $F(R)$ as a function of $R$. 
The complicated problem of going from $R_{sys}(t)$ to $P(R)$, the probability that the extension is between $R$ and $R + dR$ by accounting for fluctuations of the semi-flexible polymer handles and bead motions has been  solved using a number of {\it ad hoc} \cite{Woodside06PNAS} and precise theoretical methods \cite{Hinczewski12PNAS}. 
Assuming that $P(R)$ can be extracted from $R_{sys}(t)$ the free energy profile can be computed using $F(R) = - k_BT ln P(R)$.  
For approximate two-state systems, as is the case in P5GA hairpin \cite{HyeonBJ07,Hyeon08PNAS} or DNA hairpins  \cite{Block06Science,Woodside08COCB}, 
$F(R)$ has two dominant  minima separated by a single barrier located at $R=R_{TS}$ (Fig.~3).   
At the transition mid-force $f=f_m$, the probability of residing in the two basins of attraction should be identical (Fig.~3a), implying
$\int^{R_{TS}}_{0}e^{-\beta F_m(R)}dR=\int^{\infty}_{R_{TS}}e^{-\beta F_m(R)}dR$. 
Alternatively, $f_m$ can also be measured by equating the average dwell times in the NBA and UBA $\tau_F=\tau_U$ \cite{Bustamante01Sci,TinocoBJ06,Li07PNAS}. 
Accurate $F(R)$ profiles give estimates of the free energy barrier, $\Delta F^{\ddagger}$ and $R_{TS}$ both of which are functions of $f$. The accuracy of these estimates depend on the assumption that $R$ is a good reaction coordinate and that no information is lost in converting the measured folding trajectories ($R_{sys}(t)$ as a function of $t$) to $F(R)$.

There are two limitations that prevent extraction of the complete shape of $F(R)$. First, the probability $P(R_{TS})$ of reaching the transition state is small, making it difficult to obtain data in the neighborhood of $R_{TS}$. Second, the inferred profiles hide the possibility that there is roughness (on the length scale corresponding to base pair rupture) superimposed on  the smooth $F(R)$. These limitations were recently overcome in ingenious experiments on DNA hairpins by La Porta and coworkers \cite{LaPorta11BJ} who used a harmonic constraint to restrict $R$ to arbitrary values for long enough times to collect excellent statistics so that reliable estimates of $P(R)$ could be made (Fig.~4a). This method, which is an experimental realization of the popular umbrella sampling used in computer simulations to obtain potentials of mean force, revealed fine structure in $F(R)$ for DNA hairpins.  The superimposed fine structure on previously inferred smooth profiles perhaps reflects the rupture of base pairs (Fig.~4a), which manifests itself as "roughness" in the folding landscape.  More importantly, this study showed that $R_{TS}$ and the width of the transition region (see below on the potential relevance of $R_{TS}$)  at a given $f$ can be directly inferred from measurement.  It would be of great interest to apply this unique experimental method to study RNA molecules with tertiary interactions. 
 \\

{\bf Hopping rates and free energy profiles.}
What is the utility of $F(R)$ for biomolecules if the goal is obtain the thermodynamics and kinetics at zero or low forces in the absence of handles and beads?   Although this question has not been fully answered several groups routinely use the measured $F(R)$ and the hopping kinetics (between the folded and unfolded states in the case of hairpin) at finite $f$ to extract rates at $f = 0$ as well as the associated barrier heights ($\Delta F^{\ddagger}$)s by assuming that $R$ is an excellent reaction coordinate \cite{Block06Science,Dudko06PRL}. Because independent measurements of the absolute values of the $\Delta F^{\ddagger}$ are difficult to make the reliability of the extracted values cannot be easily assessed. 
Two computational studies, using RNA hairpin and riboswitches as illustrations, have shown the potential utility of one dimensional folding landscapes in obtaining accurate rates over a narrow range of forces close to $f_m$ \cite{HyeonBJ07,Lin08JACS}. 
In these examples, the intrinsic rates can be independently calculated using the trajectories generated in the full dimensional landscape, thus allowing for a quantitative comparison with results obtained from the projected $F(R)$.  It is now firmly established  that accurate $F(R)$ can be obtained by attaching handles that are stiff \cite{Hyeon06BJ,Hyeon08PNAS,Block06Science}, implying that the ratio $L/l_p$ ($L$ and $l_p$ are the contour length and the persistence length of the handles, respectively) should be as small as possible.  If $F(R,f)$ at $f = f_m$ is known accurately then the profiles  at arbitrary values of $f$ may be obtained using the Zhurkov-Bell relation \cite{Zhurkov65IJFM,Bell78SCI} $F(R,f) = F(R,f_m) - (f - f_m)R$.  
If $R$ is a good reaction coordinate (all other coordinates have equilibrated on time scales less that the hopping times so that the slow dynamics occurs on $F(R)$) then the hairpin formation time can be calculated using standard mean first passage time formalism,
\begin{equation}
k(f)^{-1}=\frac{1}{D_{U \rightarrow F}}\int^{R_U}_{R_F}dx e^{\beta F(x)}\int^{\infty}_xdy e^{-\beta F(y)}
\end{equation}
provided that the diffusion coefficient for transition from $U \rightarrow F$  is known.  For the P5GA hairpin, we showed that this method gives reliable results  for hopping rates over a range of $f$ around $f_m$ provided that the diffusion coefficient is calibrated by equating the theoretically calculated time at $f_m$ to the simulated value \cite{Hyeon08PNAS}. 
It might be tempting to use our method for obtaining rates at $f=0$ but this would not be  justified {\it a priori}.
\\

{\bf Molecular tensegrity and the transition state.} 
Another parameter that is extracted from $F(R)$ or suitable fits to $f$-dependent hopping rates is the location of the transition state, $R_{TS}$, which in principle moves as $f$  changes \cite{Hyeon06BJ}. 
For  $R_{TS}$, associated  with  the barrier top of $F(R)$ at $f=f_m$,  to be considered the "true" transition state, it is necessary to ensure that it is consistent  with other conventional definitions of the transition state ensemble. A plausible definition of the TS is that  the forward (to the unfolded state, $P_{unfold}$) and backward  (to the folded state, $P_{fold}$) fluxes  starting from the transition state on the reaction coordinate should be identical \cite{Klosek91BBPC}. For the hairpin it means that if an ensemble of structures were created starting at $R_{TS}$ then the dynamics in the full multidimensional space would result in these structures reaching the folded and unfolded states with equal probability. 
The number of events reaching $R_F$ and $R_U$ starting from  $R_{TS}$ can be directly  counted if folding trajectories with high temporal resolution exhibiting multiple folding and unfolding transitions at $f=f_m$ can be generated (Fig.~3a). Our coarse-grained simulations, which are the first to assess the goodness of $R_{TS}$ as a descriptor of the TS, showed that 
starting from $R_{TS}$ the hairpin crosses the TS region multiple times before reaching $R=R_F$ or $R_U$, suggesting that the TS region is broad and heterogeneous.
The transition dynamics of biopolymers occurs on a bumpy folding landscape with fine structure even in the TS region, which implies  there is an internal coordinate determining the fate of trajectory projected onto the $R$-coordinate. 
In accord with this inference we showed that for the P5GA hairpin the TS structural ensemble is heterogeneous (Fig.~3b). More pertinently,  the forward and backward fluxes starting from the structure in TS ensemble (see the dynamics of trajectories starting from $R_{TS}$ in Fig.~3b) do not satisfy the equal flux condition, $P_{fold}=P_{unfold}=0.5$. Thus, from a strict perspective $R$ for a simple hairpin may not be a good reaction coordinate, even under tension implying that $R$ is unlikely to be an appropriate reaction coordinate at $f \approx 0$. 
 
Based on simulations Morrison \emph{et al.} \cite{Morrison11PRL} proposed a fairly general theoretical  criterion to determine if $R$ could be a suitable reaction coordinate. 
The theory uses the concept of tensegrity (tensional integrity), which was introduced by Fuller and developed in the context of biology to describe stability of networks. The notion of  tensegrity has been used to account for cellular structures \cite{IngberJCS03} and more recently for the stability of globular proteins \cite{Edwards12PLoSCompBiol}, the latter of which made an interesting estimate that the magnitude of inter-residue precompression and pretension, associated with structural integrity, can be as large as a few 100 pN.         
Using $F(R)$ the experimentally measurable molecular tensegrity parameter is defined $s\equiv f_c/f_m=\Delta F^{\ddagger}(f_m)/f_m\Delta R^{\ddagger}(f_m)$, where $\Delta F^{\ddagger}=F(R_{TS})-F(R_F)$ and $\Delta R^{\ddagger}=R_{TS}-R_F$. 
The molecular tensegrity parameter $s$ represents a balance between the compression force ($f_m$) and the tensile force ($f_c$), a building principle in tensegrity systems \cite{Fuller61}.  For hairpins such stabilizing interactions are favorable base pair formation.
 In terms of $s$ and the parameters characterizing the one dimensional landscape ($f_m$ and $k_u$ in Fig. 3a) an analytic expression for $P_{unfold}$ has been obtained \cite{Morrison11PRL}.
%\begin{equation}\label{1}
%%\qquad P_{unfold}(s) = \begin{cases} 
%\frac{1}{2}\frac{1}{1+s} & s\gg 1\\ \frac{1}{2}\left[\frac{\Phi}{1+\Phi}+\frac{32f_m^2(\Phi-1)}{\pi k_u(\Phi+1)^3}s\right]^{-1} & s\ll 1 
%\end{cases}
%\end{equation}
For $R$ to be a good reaction coordinate, it is required that $P_{unfold}\approx \frac{1}{2}$. The theory has been applied to hairpins and multi-state proteins.  
Using experimentally determinable values for  $s$, one can assess whether  $R$ is a good reaction coordinate  by calculating $P_{unfold}$ using theory.  Applications of the theory to DNA hairpins \cite{Morrison11PRL}  show (Fig. 4b) that the precise sequence determines whether $R$ can be reliably used as an appropriate reaction coordinate, thus establishing the usefulness of the molecular tensegrity parameter.
 \\

{\bf Riboswitches under tension.} Riboswitches are noncoding RNA elements that sense cellular signals and 
regulate gene expression by binding target metabolites \cite{Winkler05ARM,
Breaker08Science2}. 
They  contain a conserved aptamer domain, which can bind the metabolite, and a
downstream expression platform that controls transcription termination or 
translation initiation (Fig.~5a). Riboswitches are involved in the control of both transcription and translation. Transcription termination occurs
when the downstream expression platform forms a hairpin, or a terminator stem,
followed by multi-U sequence, which results in weak interactions in the
RNA-DNA hybrid, leading to disengagement of the polymerase from the DNA template.
For riboswitches controlling translation initiation, the downstream hairpin
stem typically contains the Shine-Dalgarno sequence, the binding site
of the ribosomal unit (Fig.~5b). Formation of the
downstream hairpin, which involves a switch in the conformation of the aptamer domain (compare "ON" and "OFF" schematics in Fig.~5a), inhibits  binding of ribosomal unit needed for  
translation initiation.

Formation or disruption of the downstream 
stem, which depends on whether metabolite is bound or not, serves as a switch that controls gene expression. For riboswitches controlling transcription termination, the time needed for
metabolite binding  is typically on the order of 
seconds. For example, in purine-sensing riboswitches \cite{Edwards07COSB}, 
one of the smallest riboswitches, there are  (40-80) nucleotides between the aptamer domain and the 
multi-U sequence. With a typical transcription speed of (25-40) nucleotides 
per second, the time window is only (1-3) s for the metabolite to bind and 
stabilize the folded aptamer domain in order to influence transcription. 
On this time scale,
riboswitches, with slow metabolite binding rates, would not reach 
thermodynamic equilibrium before the terminator sequence is transcribed. Hence, in this case gene expression is under kinetic control. 

For riboswitches controlling translation initiation, folding influenced by
metabolite binding has to compete with the binding of the ribosomal unit to 
the expression platform (Fig.~5a). The rate of switching between different folding
patterns must be faster than the binding time of the ribosomal unit for the
riboswitch to function. In such riboswitches, gene expression could be under thermodynamic control since the translation initiation can occur after transcription is completed. The stability of the aptamer domain of
riboswitches may have evolved to accommodate diverse demands imposed by distinct aspects of gene expression. 

Two different but structurally identical adenine-sensing riboswitches (Fig.~5b),
{\it pbuE} and {\it add} adenine (A) riboswitches, that have been studied using pulling experiments \cite{Greenleaf08Science} and simulations \cite{Lin08JACS} show similar aptamer 
structure but control gene expression through different functions 
\cite{Serganov04JCB,Mandal03Cell,Mandal04NSMB}. Both 
{\it pbuE} and {\it add} A-riboswitches are on-switch riboswitches, but 
{\it pbuE} A-riboswitch controls transcription termination while {\it add} 
A-riboswitch controls translation initiation. In order to describe the functions of riboswitches  it is important to understand the time scales associated with the folding of the aptamer, and the conformational changes involving part of the aptamer that forms the terminator hairpin by base pairing with the transcript in the downstream expression platform. In other words, one needs to understand in quantitative terms the folding landscape of the riboswitches in the presence and absence of metabolites so that the transition rates to specific states can be obtained. Such detailed information is best obtained using single-molecule methods complemented by suitable theory. 
\\

{\bf Stability of helices determine $f$-induced unfolding of purine 
riboswitches.} 
The adenine riboswitch aptamer is a 
three-way junction formed from helix P1 and hairpins P2 and P3 (Fig.~5b). The junction
contains the binding pocket for adenine (Fig.~5b), which is stabilized by tertiary interactions in the folded
state. There is a kissing loop-loop interaction
between helices P2 and P3 (Fig.~5b), which is transient in the 
absence of metabolites, and is stabilized only when  the aptamer binds the ligand.

A single-molecule experiment first observed the 
hierarchical folding in the {\it pbuE} adenine riboswitch aptamers 
\cite{Greenleaf08Science} by measuring the extension of the riboswitch aptamer as a function of $f$.
In the absence of the metabolite, two clear steps in the change of contour 
lengths were found, which were associated with the unfolding of helices P2
and P3. Because of the difference in the lengths of the helices P2 and P3
(19 and 21 nucleotides, respectively), the order of unfolding of the two helices
could be unambiguously assigned.   The measured folding landscape shows (Fig.~6a) that  helix P3 is ruptured  before P2 in the 
{\it pbuE} A-riboswitch aptamer. With metabolite bound, larger changes in
contour lengths with larger forces were found, and unfolding occurred in a single step.
This is due to the stabilization of the folded aptamer structure by the bound metabolite, which makes unfolding of helix P1 and the binding pocket the major free energy barrier in the unfolding process. Fig.~6a shows that refolding from an initially  unfolded state occurs 
 with the dominant pathway being
order of $P2\rightarrow P3 \rightarrow P1$. 

A theoretical study of the {\it add} adenine riboswitch 
aptamer using a coarse-grained self-organized polymer model also found
multiple intermediate steps in the folding landscape of the riboswitch 
aptamer \cite{Lin08JACS}. In order to ascertain the folding mechanism, simulations were performed by 
quenching the force from an initial high force to constant low forces. 
%The forces at which P3, P2, and P1 form are only modestly larger than the 
%experimental values for the {\it pbuE} riboswitch.
During the folding process P3 forms first followed by P2, and finally the
triple-helix junction and the helix P1 form. The order of folding can be 
directly associated with the stability of individual helices \cite{Lin08JACS}. Indeed, the 
isolated P3 is more stable than P2 by about 1 kcal/mol as predicted by the 
Vienna RNA package \cite{Hofacker03NAR}, thus explaining its early formation. 

Remarkably, despite the structural similarity 
between {\it pbuE} and {\it add} A-riboswitch aptamers,
experiments show that P2 in {\it pbuE} unfolds last (Fig.~6a) and presumably is
the first structural element to refold. The organization principle that emerges is that assembly of RNA is 
largely determined by the stability of individual helices, which  implies that in the 
{\it pbuE} riboswitch aptamer, P2 ought to be more stable than P3.
Indeed, the relative stability of P2/P3 is different in the {\it pbuE}
A-riboswitch aptamer.  The predicted free energy for the secondary
structure of P2 is smaller than that of P3 by 2 kcal/mol. The stability
difference explains the reversed order of the folding of P2 and P3 shown in 
the folding landscapes.

Our prediction for the folding landscape of {\it add} adenine riboswitch (Fig.~6b) was quantitatively validated in a recent single-molecule 
experiment on {\it add} A-riboswitch \cite{Neupane11NAR}.
The  experiments showed that helix P1 in {\it add} A-riboswitches is 
more stable than that in {\it pbuE} A-riboswitches. In {\it pbuE} 
A-riboswitches, helix P1 unfolds at forces as low as $\sim 3$ pN 
without metabolite being bound. \cite{Greenleaf08Science}.
In {\it add} A-riboswitches, forces $\sim 10$ pN are needed to unfold helix P1 
\cite{Lin08JACS,Neupane11NAR}. The folding landscapes of these two purine 
riboswitches, obtained from experiments and simulations, show that the assembly of riboswitches
is determined by the local stability of the structural 
elements in these two purine riboswitch aptamers.
The principle linking folding mechanism to stability of individual helices is general and has been further illustrated using folding of  a number of 
psuedoknots \cite{Cho09PNAS} whose folding pathways could be accurately predicted using the 
stability of the individual helices.  It might appear that entropic effects could become relevant when folding takes place spontaneously (in the absence of force), thus invalidating the proposed organization principle. This is not the case as  simulations have explicitly demonstrated \cite{Cho09PNAS}.  More importantly, the organization principle described here has been used to quantitatively rationalize temperature jump experiments on RNA pseudoknots \cite{narayanan2011JACS}.\\

{\bf Transition rates between the network of connected states from {\it add} 
A-riboswitch}. Just as for RNA hairpins, the matrix of transition rates 
connecting the various states (F, P2/P3, P3, and U in Fig.~6c) 
could be calculated using the free energy profile in Fig.~6b. The $f$-dependent transition rate between 
any two states can be obtained from the time
traces of the extension, which would require generating a large number of folding 
trajectories at each $f$.   Alternatively, $F(f, R)$ calculated from 
accurate estimates at one force could be used to obtain the rates at other forces based upon 
the theory of mean first passage times (Eq.~1). Following the procedure used for 
obtaining hopping rates for hairpins away from $f_m$ \cite{Hyeon08PNAS}, we calculated the 
diffusion coefficients (needed to obtain hopping rates between multiple states of the riboswitch) by equating the transition rate calculated using 
Kramers theory to that obtained from time traces generated using simulations at a specified $f$.
% = 12 pN for 
%$U\rightarrow P3$ and $P3\rightarrow P2/P3$ transition and at f = 10 pN for 
%$F\rightarrow P2/P3$ transition. 
These effective diffusion coefficients reflect collective processes associated 
with global folding of the riboswitch. Thus, we expect that generically $D \approx \sigma^2/\tau_0$ 
where $\tau_0$ ($\approx 10^{-6}$ sec) the prefactor in RNA folding 
\cite{Thirum05Biochem,Hyeon12BJ}. If $\sigma = 0.7 nm$, then the calculated 
value of $D \approx 10^4 nm^2/s$, which is in reasonable agreement with the 
numerical values needed for obtaining agreement between rates from 
simulations and from free energy profiles \cite{Lin08JACS}.
%The resulting diffusion coefficients obtained are 
%$D_{U \leftrightarrow P3}$ = 4.3 $\times10^4$ $nm^2/s$, 
%$D_{P3\leftrightarrow P2/P3}$ = 8.7$\times 10^4$ $nm^2/s$, and
%$D_{F \leftrightarrow P2/P3}$ = 2$\times 10^5$ $nm^2/s$.  
Using the values of $D$ the hopping rates between the various states in {\it add} A-riboswitch were calculated over a range of forces using Eq. (1) with $F(R,f)$ at one value of $f$! Comparison of rates obtained using $F(R,f)$ and Eq. (1) and numerically exact results obtained from simulations (Fg.~6c) is excellent.    This might suggest that $R$ may be a reasonable reaction coordinate for describing riboswitch folding. However, when we assess if the location of the barrier height describes the actual transition state ensemble for the $F\rightarrow P2/P3$ transition using the molecular tensegrity parameter it is found that $P_{fold}$ deviates substantially from 0.5. Thus, it is unlikely that for complex structures such as riboswitches, and more generally ribozymes, the pulling coordinate alone can describe the folding process. \\

{\bf Summary.} Although fundamentally new insights into the folding of RNA have emerged from subjecting them to tension there are many major unresolved questions, which will require a combination of theory, simulations and experiments. Here are a few of them. (i) In the most favorable circumstances the folding landscape and hopping rates can only be obtained at force values at which RNA makes multiple transitions. Can these estimates be used to extrapolate to zero force? (ii) Applications here demonstrate that extension is not always a good reaction coordinate. If this is the case how can one profitably utilize the high quality folding trajectories to map the network of connected states, which could be hidden in $R$ but become transparent in other auxiliary variables? (iii) Even if $R$ is a good reaction coordinate it is unclear if the extracted values for mean barrier heights at zero force are reliable? This issue is exacerbated because in general one expects a distribution of barrier heights for RNA and proteins \cite{Klimov01JPCB}, and hence the mean value may not be informative. In addition, independent measurements of the absolute values of barrier heights are difficult thus making it hard to assess the accuracy of estimates based on single-molecule folding trajectories. Only by resolving these issues  the full scope  and power of single-molecule pulling experiments in biology can be realized.    
\\

\textbf{Acknowledgments:} We thank the National Science Foundation (CHE 09-4033) and the National Institutes of Health (GM089685) for supporting this work.

%\bibliographystyle{pnas2010}
%\bibliography{mybibNew}

\clearpage

\begin {figure}
 \includegraphics[width=6.20in]{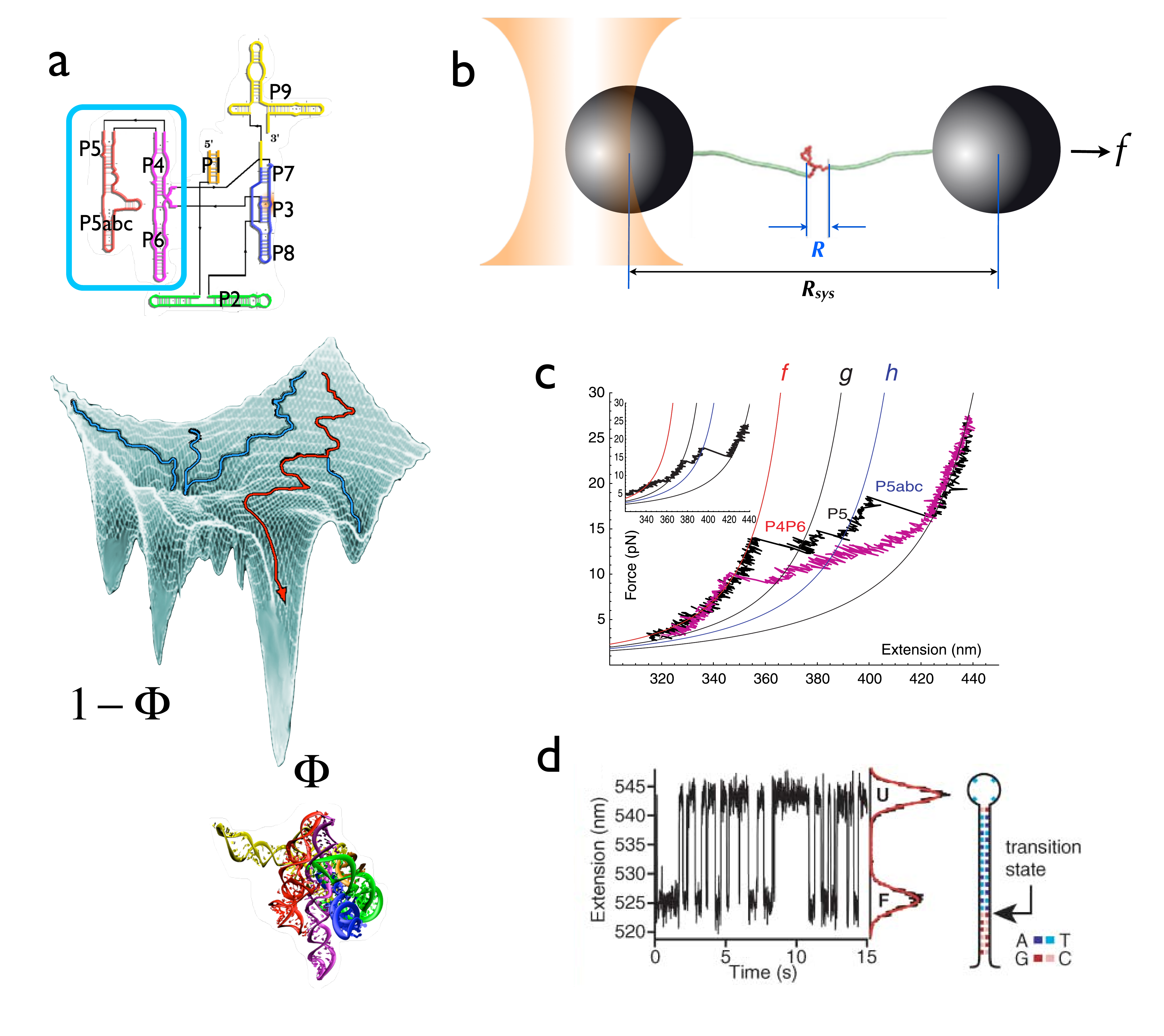}
 \caption{
{\bf a.} A schematic of folding landscape and kinetic partitioning. Secondary structure map and folded states of {\it Tetrahymena} ribozyme are depicted on the top and bottom of the landscape cartoon. The P4-P6 domain is highlighted in the cyan box.
 {\bf b.} Illustration of the setup of a laser optical tweezers experiment for a RNA hairpin. 
The figure illustrates that the end-to-end distance ($R$) dynamics of RNA hairpin is indirectly monitored through the dynamics of distance between the two microbeads ($R_{sys}$).  
 {\bf c.} Force extension curve of P4-P6 domain of {\it Tetrahymena} ribozyme. Order of unfolding of the structural elements in the P4-P6 domain are extracted from the measured FECs as indicated.  {\bf d.} Histogram of end-to-end distance (extension) ($P(R)$) of a DNA hairpin (shown on the right) are obtained directly from the time trace of $R_{sys}(t)$.   Figures in c, d are adapted from Ref. \cite{Bustamante03Science,Block06Science}
}
 \end{figure}\clearpage 
 
\begin{figure}
 \includegraphics[width=6.20in]{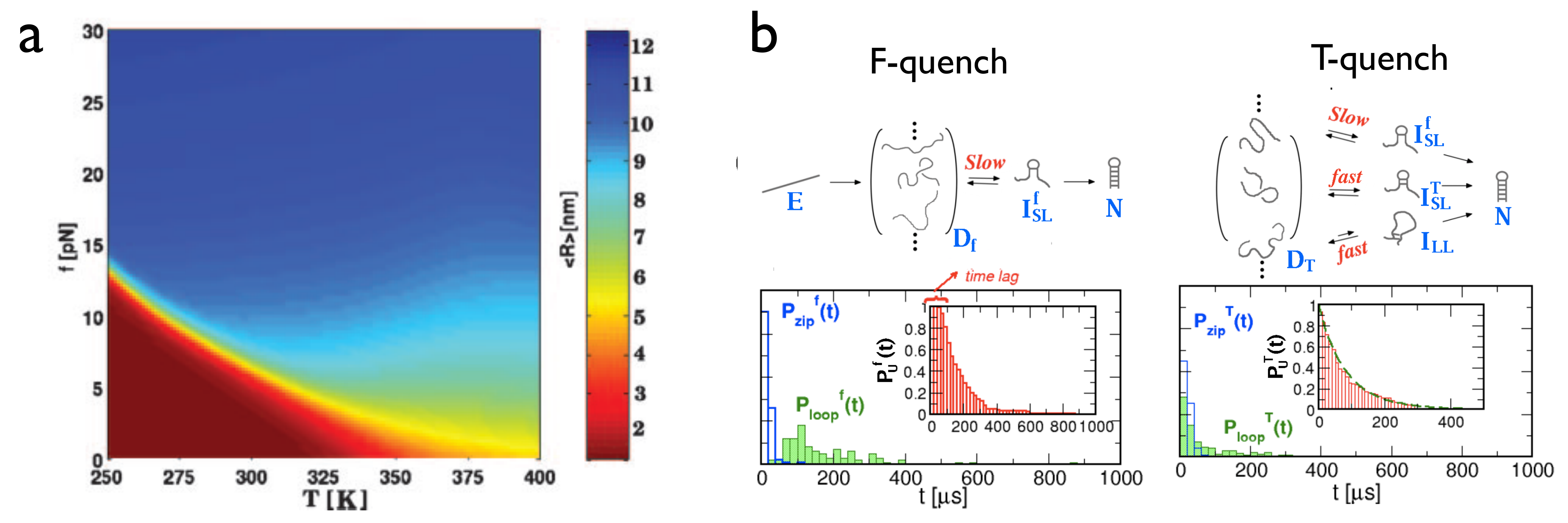}
 \caption{
 {\bf a.} ($T$,$f$) phase diagram of P5GA hairpin using $R$ (scale on the right) as an order parameter. Blue is unfolded and dark red corresponds to the hairpin state.
{\bf b.}  Demonstration that the folding mechanisms of P5GA hairpin vary depending on the protocol used to initiate folding. On the left we show the schematics of approach to the hairpin state using force-quench and the right shows the events upon temperature-quench.  The bottom graphs show 
decompositions of folding time ($\tau_{FP}$) under these protocols two into times for looping ($\tau_{loop}$) and zipping time ($\tau_{zip}$). 
While the distributions of $\tau_{zip}$ are similar for both conditions, 
the $\tau_{loop}$ upon $f$-quench is longer  than under $T$-quench condition and is more broadly distributed.    
The fraction of unfolded molecules at each condition ($P_U^f(t)$ and $P_U^T(t)$) is plotted in the inset. 
$P_U^f(t)$ that is fit to $P_U^f(t) = e^{-(t - 50\mu s)/138 \mu s}$ for 
$t>50\mu s$ shows a lag phase at $0<t<50$ $\mu s$ 
suggesting that the state $I^f_{SL}$ is an obligatory step for the refolding process under $f$-quench.  
In contrast, $P_U^T(t)$ is well fit using a sum of two exponential functions $P_U^T(t) = 0.44\times e^{-t/63 \mu s} + 0.56\times e^{-t/104 \mu s}$.
}
\end{figure}

\clearpage

\begin {figure}
 \includegraphics[width=6.20in]{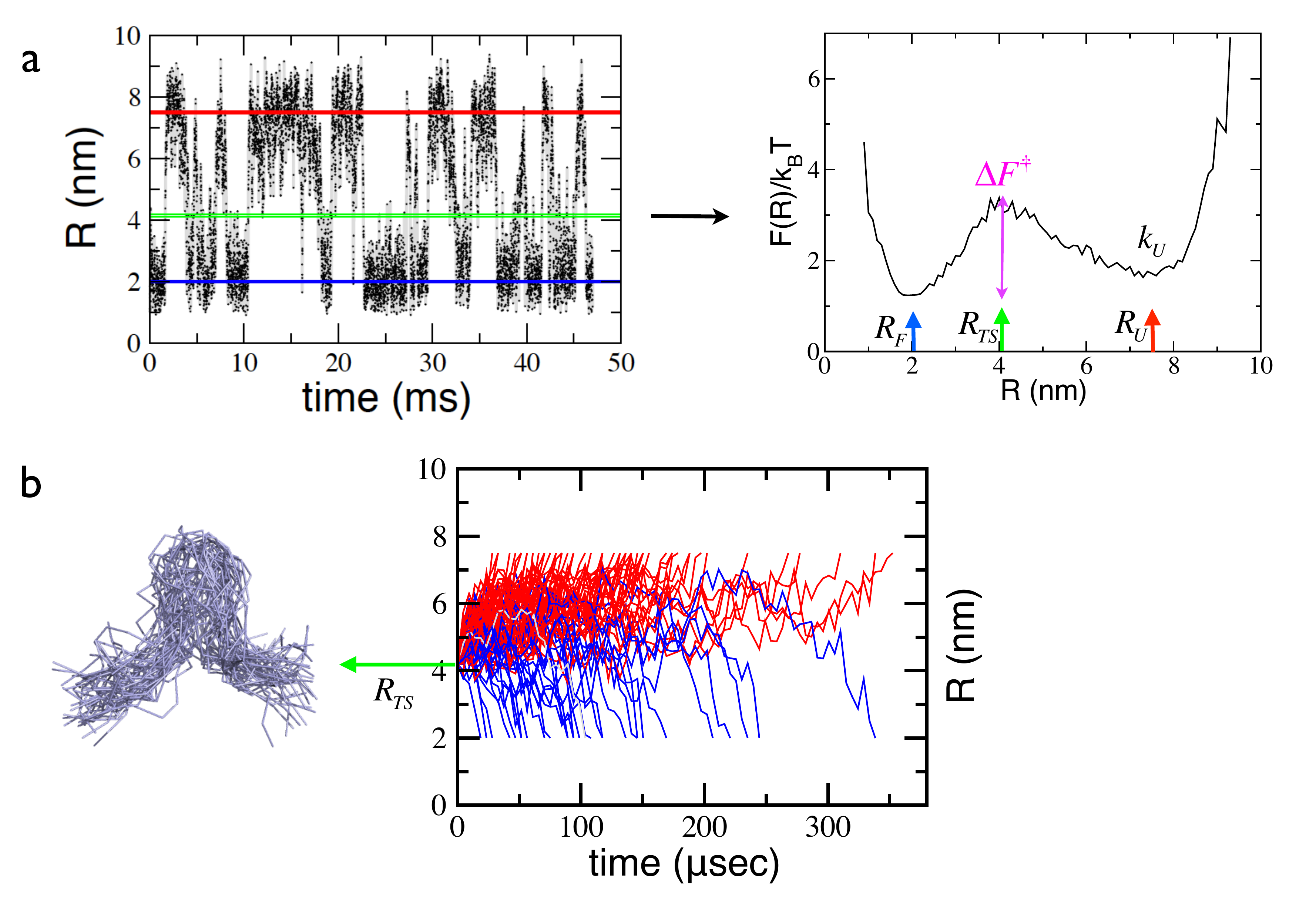}
 \caption{
{\bf a.} Time trace of end-to-end distance ($R$) of  RNA hairpin at transition mid-force $f_m=14.7$ pN (left). The  corresponding free energy profile in terms of $R$, $F(R)$ (right). The positions of native, unfolded, and transition states are marked with arrows. 
In addition, barrier height ($\Delta F^{\ddagger}$) and the curvature of unfolded state ($k_U$) are also shown on the $F(R)$. 
{\bf b.} Time trajectories of simulations starting from the configurations of transition state ensemble (shown on the left). Trajectories reaching the folded and unfolded state at 2.0 nm and 7.5 nm are colored in blue and red, respectively. In the blue trajectories, a number of recrossing events can be observed. The figures are adapted from Ref. \cite{Hyeon05PNAS,Hyeon08JACS}
}
\end{figure}
\clearpage

\begin{figure}
 \includegraphics[width=6.20in]{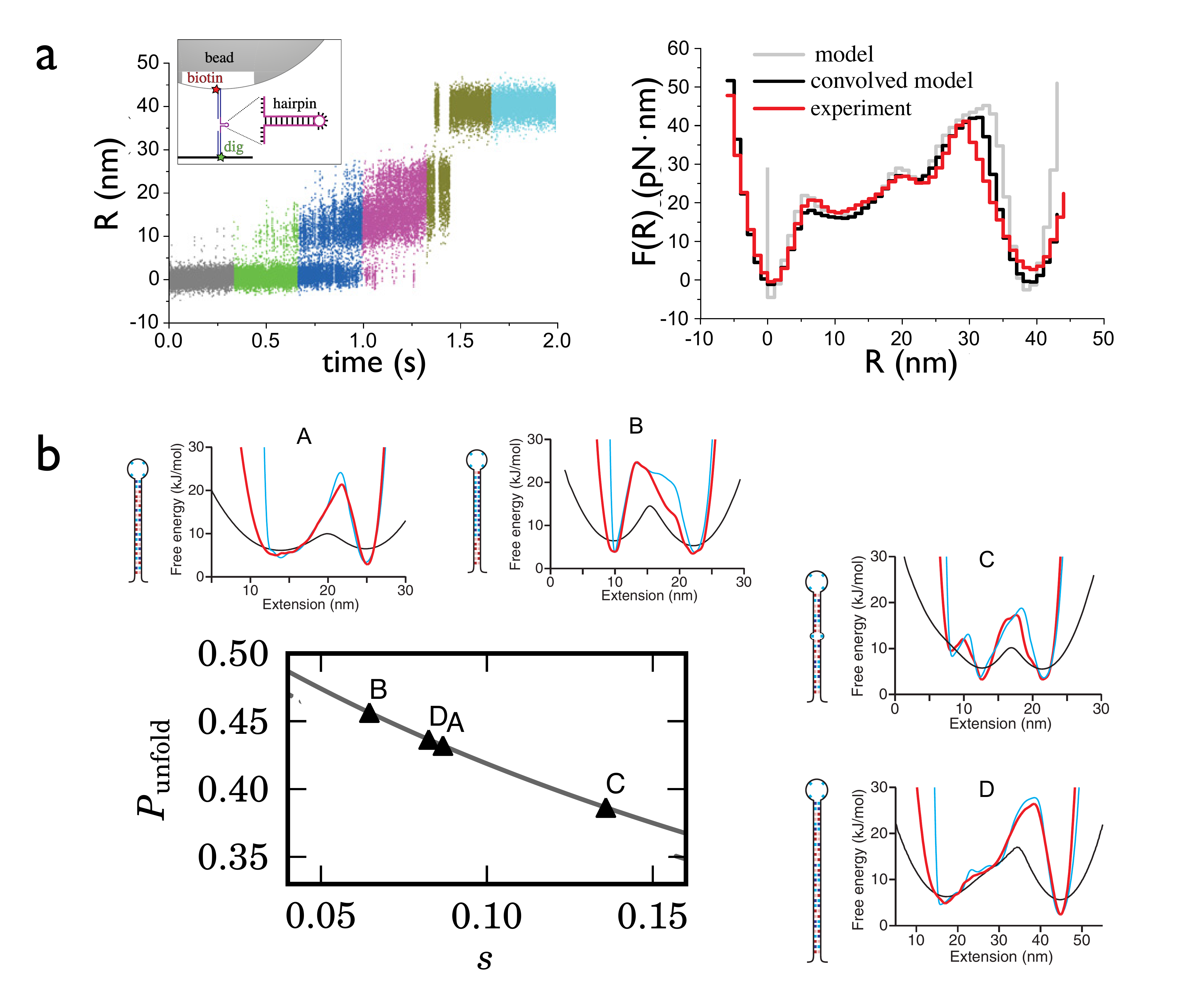}
 \caption{
{\bf a.} End-to-end distance dynamics of DNA hairpin under gradually increasing harmonic constraint. Gradual change of $R$-dynamics are depicted in the folding trajectory trajectory.  
Free energy profile reconstructed by using the harmonic constraining  method (umbrella sampling) at the transition mid-force is shown on the right.
{\bf b.} Tensegrity parameters calculated for four DNA hairpins with different sequences in Ref.\cite{Block06Science} is related to $P_{unfold}$. 
The DNA hairpin with sequence B is predicted to have $P_{unfold}$ most proximal to 0.5, which suggests that the free energy profile calculated in terms of end-to-end distance coordinate most accurately describe the dynamics of this DNA hairpin. 
The figure is adapted using Ref.\cite{Block06Science,Morrison11PRL}
}
\end{figure}
\clearpage

\begin {figure}
 \includegraphics[width=6.20in]{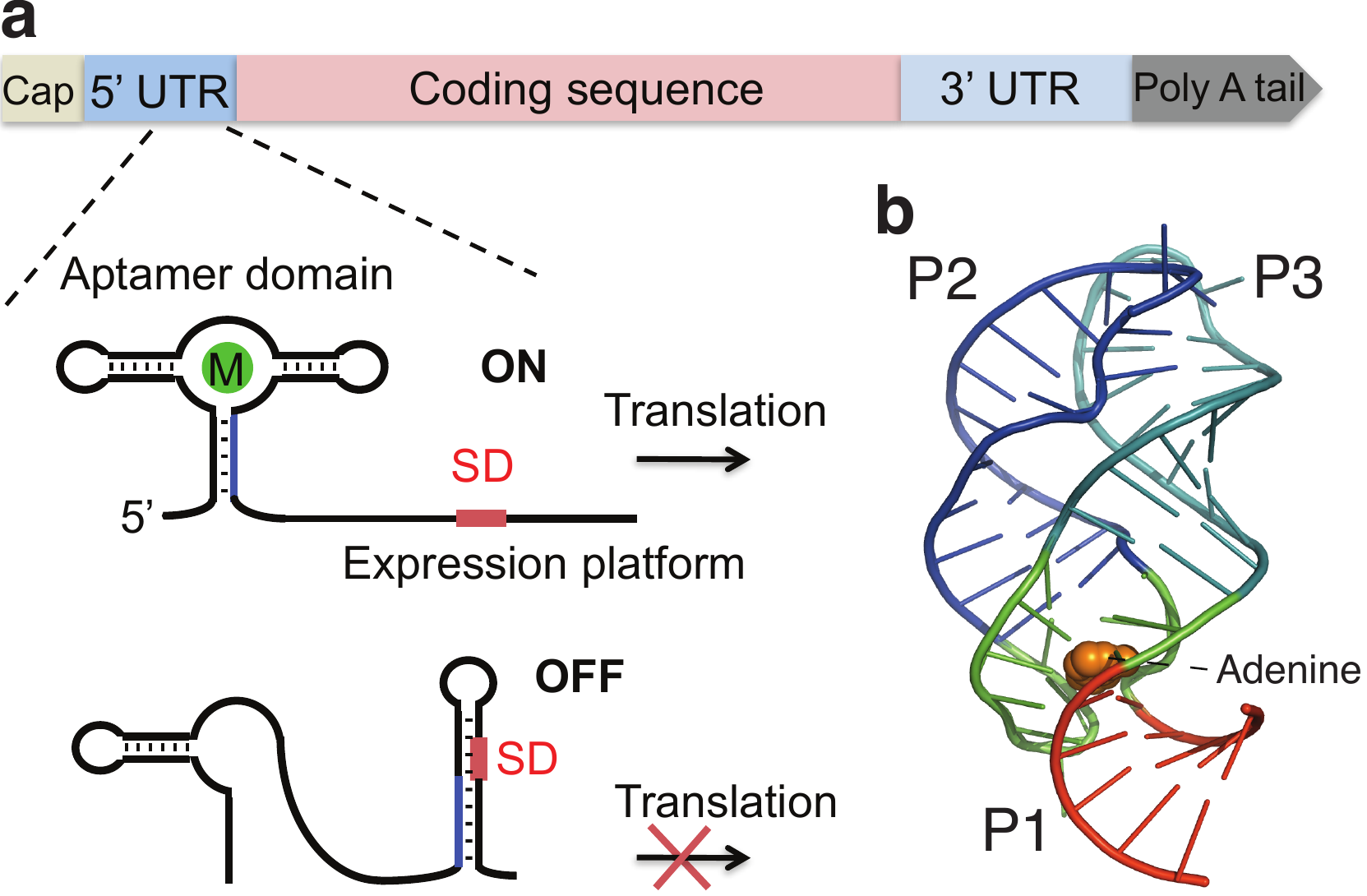}
 \caption{{\bf a}. Schematic of a riboswitch, which is in the 5' untranslated region (5'UTR) in mRNA. The {\it add} adenine riboswitch (blow up of the 5'UTR given below) contains an aptamer domain,
which can bind the metabolite, adenine, and a downstream expression
platform, which contains the Shine-Dalgarno sequence (SD). When  adenine
is bound to the aptamer domain, the SD  binds to the  ribosomal unit for
translation initiation denoted by ON. When the riboswitch does not bind adenine, the SD forms
 a hairpin with the downstream expression platform (denoted by OFF), which prevents the ribosomal unit from recognizing the SD, thus resulting in translation being inhibited.   {\bf b.} Tertiary structure
of the aptamer domain of the {\it add} A-riboswitch with adenine (shown in orange) bound. The 
aptamer form a three-way junction comprised of  helix P1 and hairpins P2 and P3.
Adenine binding stabilizes the tertiary loop-loop interactions between
P2 and P3.
}
\end{figure}
\clearpage

\begin {figure}
 \includegraphics[width=4.0in]{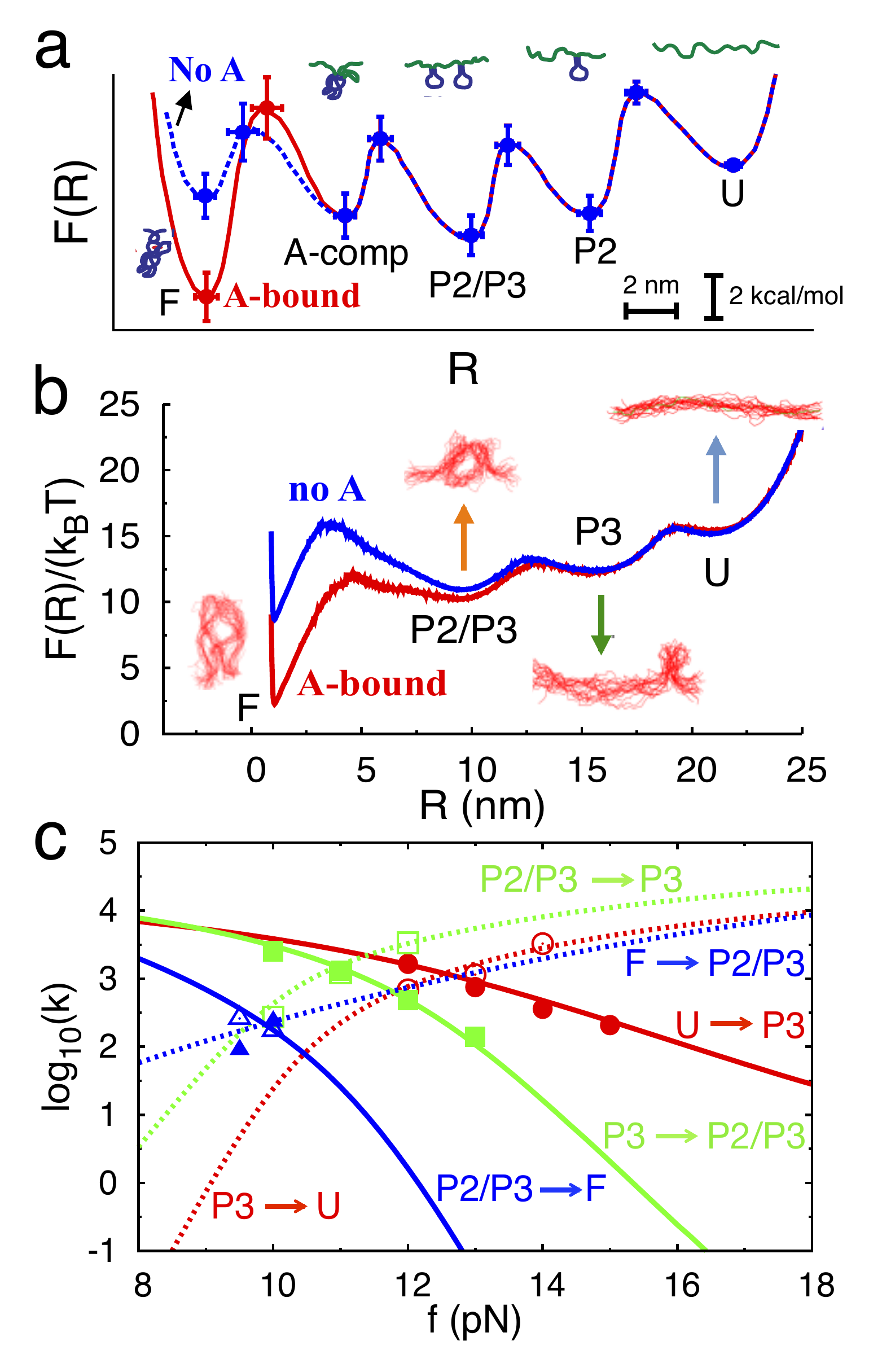}
 \caption{
{\bf a.} Free energy profile extracted and reconstructed from LOT experiments for {\it pbuE} A-riboswitch with and without adenine being bound at $f = 6.5$ pN \cite{Block06Science}. The structural elements (see Fig.~5b) that are intact in each state are indicated. {\bf b.} Folding landscape for \emph{add} A-riboswitch aptamer as a function of 
extension R at $f = 10$ pN without and with adenine. The ensemble of structures
in the intermediate states are shown. {\bf c.} The logarithm of the transition rates,
log($k=1/\bar{\tau}$), between the distinct states calculated using the theory of mean first passage times
 (Eq.1, shown in lines), and directly from the time traces 
of the extension of a coarse-grained model of the aptamer generated using Brownian dynamics simulations (shown as points).
}
\end{figure}

\end{document}